\documentclass[11pt]{article}
\usepackage{moriond,epsfig}

\def \gh{\vphantom{$\fbox{\Big[}$}}

\def \beq{\begin{equation}}
\def \eeq{\end{equation}}
\def \bea{\begin{eqnarray}}
\def \eea{\end{eqnarray}}
\def \ben{\begin{enumerate}}
\def \een{\end{enumerate}}
\def \bit{\begin{itemize}}
\def \eit{\end{itemize}}

\def \branch{{\cal B}}
\def \eff{\hbox{eff}}
\def \gev{{\hbox{GeV}}}

\def \tev{{\hbox{TeV}}}

\def \cl#1{{#1\%\ \mathrm{C.L.}}}

\def \eq#1{Eq.~(\ref{#1})}

\def \fig#1{Fig.~\ref{#1}}

\def \rf{Ref.~\cite}

\def \b{\beta}

\def \g{\gamma}

\def \d{\delta}

\def \m{\mu}

\begin{document}
\vspace*{4cm}
\title{NNLO ANALYSIS OF SEMILEPTONIC RARE $B$ DECAYS}
\author{ E. LUNGHI }
\address{
Deutsches Elektronen Synchrotron, DESY, \\
Notkestrasse 85, D-22607 Hamburg, Germany}
\maketitle\abstracts{ We update the standard model predictions for the
branching ratios of the decays $B\to (K,K^*,X_s) \; \ell^+ \ell^-$,
with $\ell=e,\; \m$. Using the measurements of $\branch (B\to X_s \g)$
and $\branch (B\to K \ell^+ \ell^-)$ as well as the upper limits on
$\branch (B\to (K^*,X_s) \ell^+ \ell^-)$, we work out model
independent bounds on the relevant Wilson coefficients. We show the
impact of such bounds on the parameter space of the MSSM with minimal
flavour violation and of supersymmetric models with additional sources
of flavour changing.}
\section{Introduction}
With increased statistical power of experiments at the $B$--factories,
the exclusive and inclusive $B\to (K,K^*,X_s) \; \ell^+ \ell^-$ decays
will be measured very precisely, in the next several years. On the
theoretical side, partial results in next-to-next-to-leading
logarithmic accuracy are now available in the inclusive channels $B
\to X_s \ell^+ \ell^-$~\cite{BMU,AAGW}. In this talk, we review the
status of the standard model predictions for the inclusive and
exclusive modes and present the model independent constraints implied
by the new data (see Ref.~\cite{aghl} for a more detailed presentation
of the results). The experimental input that we use in our analysis is
given below. Except for the inclusive branching ratio for $B\to X_s
\g$, which is the average of the results from CLEO~\cite{cleobsg},
ALEPH~\cite{alephbsg} and BELLE~\cite{bellebsg} measurements, all
other entries are taken from the two BELLE papers listed in
Ref.~\cite{bellebsll}:
\bea
\branch (B\to X_s \g) &=& (3.22\pm 0.40) \times 10^{-4} \; , 
\label{bsgexp}\\
\branch (B\to K \mu^+ \mu^-) &=& (0.99^{+0.40+0.13}_{-0.32-0.14})
 \times 10^{-6} \; , \label{bkmmexp}\\
\branch (B\to K e^+ e^-) &=& (0.48^{+0.32+0.09}_{-0.24-0.11})\times
10^{-6} \; , \label{bkeeexp} \\
\branch (B\to K \ell^+ \ell^-) &=& (0.75^{+0.25}_{-0.21}\pm 0.09)\times
10^{-6} \; , \label{bkllexp} \\
\branch (B\to K^* \mu^+ \mu^-) &\leq& 3.0 \times  10^{-6} \; {\rm at}\;
 \cl{90}  \; \label{bksmmexp} , \\
\branch (B\to K^* e^+ e^-) &\leq& 5.1 \times  10^{-6} \; {\rm at}\;   
 \cl{90} \;  \label{bkseeexp} ,\\ 
\branch (B\to X_s \mu^+ \mu^-) &\leq & 19.1 \times  10^{-6} \; {\rm at}\;
 \cl{90} \;   \label{bsmmexp} ,\\
\branch (B\to X_s e^+ e^-) &\leq& 10.1 \times  10^{-6} \; {\rm
at}\; \cl{90}
 \;  . \label{bseeexp} 
\eea
The upper bounds given in Eqs.~(\ref{bkmmexp}) -- (\ref{bseeexp})
refer to the so--called non--resonant branching ratios integrated over
the entire dilepton invariant mass spectrum. In the experimental
analyses, judicious cuts are used to remove the dominant resonant
contributions arising from the decays $B \to (X_s,K,K^*)
(J/\psi,\psi^\prime,...) \to (X_s,K,K^*) \ell^+ \ell^-$. A direct
comparison of experiment and theory is, of course, very desirable, but
we do not have access to this restricted experimental
information. Instead, we compare the theoretical predictions with data
which has been corrected for the experimental acceptance using
SM-based theoretical distributions. In the present analysis, we are
assuming that the acceptance corrections have been adequately
incorporated in the experimental analysis in providing the branching
ratios and upper limits listed above. We will give the theoretical
branching ratios integrated over all dilepton invariant masses to
compare with these numbers. However, for future analyses, we emphasize
the dilepton invariant mass distribution in the low-$\hat{s}$ region,
$\hat{s} \equiv m_{\ell^+ \ell^-}^2/m_{b, pole}^2 \leq 0.25$, where
the NNLO calculations for the inclusive decays are known, and resonant
effects due to $J/\psi$, $\psi'$, etc. are expected to be small.

The effective Hamiltonian in the SM inducing the $b\to s \ell^+
\ell^-$ and $b\to s\g$ transitions is:
\begin{eqnarray}
    \label{Heff}
    {\cal H}_{\eff} =  - \frac{4G_F}{\sqrt{2}} V_{ts}^* V_{tb}
    \sum_{i=1}^{10} C_i(\mu) \, O_i (\mu)\quad ,
\end{eqnarray}
where $O_i(\mu)$ are dimension-six operators at the scale $\mu$,
$C_i(\mu)$ are the corresponding Wilson coefficients, $G_F$ is the
Fermi coupling constant, and the CKM dependence has been made
explicit.  The operators can be chosen as in \rf{BMU}
$$
\begin{array}{rclrcl}
    O_1    & = & (\bar{s}_{L}\gamma_{\mu} T^a c_{L })
                (\bar{c}_{L }\gamma^{\mu} T^a b_{L}) \, , &
    O_2    & = & (\bar{s}_{L}\gamma_{\mu}  c_{L })
                (\bar{c}_{L }\gamma^{\mu} b_{L}) \, , \\ \vspace{0.2cm}
    O_3    & = & (\bar{s}_{L}\gamma_{\mu}  b_{L })
                \sum_q (\bar{q}\gamma^{\mu}  q) \, , &
    O_4    & = & (\bar{s}_{L}\gamma_{\mu} T^a b_{L })
                \sum_q (\bar{q}\gamma^{\mu} T^a q) \, , \\ \vspace{0.2cm}
    O_5    & = & (\bar{s}_L \gamma_{\mu_1} \gamma_{\mu_2}
                \gamma_{\mu_3}b_L)
                \sum_q(\bar{q} \gamma^{\mu_1} \gamma^{\mu_2}\gamma^{\mu_3}q)
                \, , &
    O_6    & = & (\bar{s}_L \gamma_{\mu_1} \gamma_{\mu_2}
                \gamma_{\mu_3} T^a b_L)
                \sum_q(\bar{q} \gamma^{\mu_1} \gamma^{\mu_2}
                \gamma^{\mu_3} T^a q) \, , \vspace{0.2cm} \\
\vspace{0.2cm}
    O_7    & = & \frac{e}{g_s^2} m_b (\bar{s}_{L} \sigma^{\mu\nu}
                b_{R}) F_{\mu\nu} \, , &
    O_8    & = & \frac{1}{g_s} m_b (\bar{s}_{L} \sigma^{\mu\nu}
                T^a b_{R}) G_{\mu\nu}^a \, , \\ \vspace{0.2cm}
    O_9    & = & \frac{e^2}{g_s^2}(\bar{s}_L\gamma_{\mu} b_L)
                \sum_\ell(\bar{\ell}\gamma^{\mu}\ell) \, , &
    O_{10} & = & \frac{e^2}{g_s^2}(\bar{s}_L\gamma_{\mu} b_L)
                \sum_\ell(\bar{\ell}\gamma^{\mu} \gamma_{5} \ell) \, ,
\end{array}
$$
where the subscripts $L$ and $R$ refer to left- and right- handed
components of the fermion fields.
\section{Expectations in the standard model}
\begin{table}[t]
\caption{SM predictions at NNLO accuracy for the various inclusive and
exclusive decays involving the quark transition $b\to s \ell^+
\ell^-$. For the exclusive channels the indicated errors correspond to
variations of the form factors, $\mu_b$, $m_{t,pole}$ and $m_c/m_b$,
respectively. For the inclusive channels the errors correspond,
respectively, to variations of $\mu_b$, $m_{t,pole}$ and $m_c/m_b$. In
the last column we add the errors in quadrature. \hspace*{6cm}}
\vspace{0.4cm}
\begin{center}
\begin{tabular}{| l | l | l |}
\hline
$B\to K\ell^+\ell^-$&$\left(0.35\pm 0.11\pm 0.04\pm 0.02\pm 0.0005
 \right) \times 10^{-6}$ & $\left(0.35\pm  0.12 \right) \times 10^{-6}$ \gh \\ \hline
$B\to K^*e^+e^-$&$\left(1.58\pm 0.47\pm 0.12^{+0.06}_{-0.08}\pm 0.04
 \right) \times 10^{-6}$ & $\left(1.58 \pm  0.49 \right) \times 10^{-6}$  \gh \\ \hline
$B\to K^*\m^+\m^-$&$\left(1.19\pm 0.36\pm 0.12^{+0.06}_{-0.08}\pm 0.04
 \right) \times 10^{-6}$ & $\left(1.19 \pm  0.39 \right) \times 10^{-6}$  \gh \\ \hline
$B\to X_s \mu^+ \mu^-$ & $\left(4.15 \pm 0.27\pm 0.21\pm 0.62 \right)
 \times 10^{-6}$ & $\left(4.15 \pm  0.70 \right) \times 10^{-6}$  \gh \\ \hline
$B\to X_s e^+ e^-$ & $\left(6.89 \pm 0.37 \pm 0.25\pm 0.91\right)
 \times 10^{-6}$ & $\left(6.89 \pm  1.01 \right) \times 10^{-6}$  \gh \\ \hline
\end{tabular}
\end{center}
\label{table:sm}
\end{table}
The SM predictions for semileptonic rare $B$ decays are summarized in
table~\ref{table:sm}. 

Note that a genuine NNLO calculation of the inclusive branching ratios
only exists for values of $\hat s$ below 0.25.  For high values of
$\hat s$, an estimate of the NNLO result is obtained by an
extrapolation procedure. It is possible to show~\cite{aghl}, that the
full NNLO invariant mass distribution is very well approximated, in
the entire low-$\hat s$ range, by the partial NNLO\footnote{By partial
NNLO we mean that all the terms, for whom the low--$\hat s$ assumption
was computationally essential, have to be dropped (see
Ref.~\cite{aghl} for more details).}  for the choice of the scale
$\mu=2.5 \;\gev$. This is yet another illustration of the situation
often met in perturbation theory that a judicious choice of the scale
reduces the higher order corrections. It seems, therefore, reasonable
to use the partial NNLO curve corresponding to $\mu_b = 2.5 \; \gev$
as an estimate for the central value of the full NNLO for
$\hat{s}>0.25$.
 
For what concerns the exclusive decays $B\to K^{(*)} \ell^+ \ell^-$,
we implement the NNLO corrections calculated by Bobeth {\it et
al.}~\cite{BMU} and by Asatrian {\it et al.}~\cite{AAGW} for the
short-distance contribution. Then, we use the form factors calculated
with the help of the QCD sum rules in \rf{Ali:2000mm}. In order to
accommodate present data on the $B\to K^* \gamma$ decay, we use the
minimum allowed form factors, given in Table~5 of \rf{Ali:2000mm}, as
our default set. In our numerical analysis, we add a flat $\pm 15\%$
error as residual uncertainty on the form factors.

Let us stress once more that, from a theoretical point of view, would
be much better to compare the non--resonant branching ratios
integrated in the low--$\hat s$ region with the corresponding
experimental bounds. Defining such region according to the Belle
analysis presented in \rf{bellebsll}, we choose the integration limits
as follows:
\bea
B\to X_s e^+ e^-: \; \left( 0.2\; \gev \over m_b \right)^2 \leq \hat s \leq 
      \left( M_{J/\Psi} - 0.6 \; \gev \over m_b \right)^2 \, , \\
B\to X_s \m^+ \m^-: \; \left( 2 m_\mu \over m_b \right)^2 \leq \hat s \leq 
      \left( M_{J/\Psi} - 0.35 \; \gev \over m_b \right)^2 \; .
\eea
For the integrated branching ratios in the SM we find:
\bea
\branch (B\to X_s e^+ e^-) \Big|_{\rm low} &=& \left(2.47 \pm 0.40 \right) \times 10^{-6} \;\;\; 
        (\d \branch_{X_see} = \pm 16 \%)\; , \\
\branch (B\to X_s \m^+ \m^-) \Big|_{\rm low} &=& \left(2.75 \pm 0.45 \right) \times 10^{-6} \;\;\; 
        (\d \branch_{X_s\m\m} = \pm 16 \%) \; . 
\eea
\section{Model independent analysis}
\begin{figure}
\begin{center}
\epsfig{file=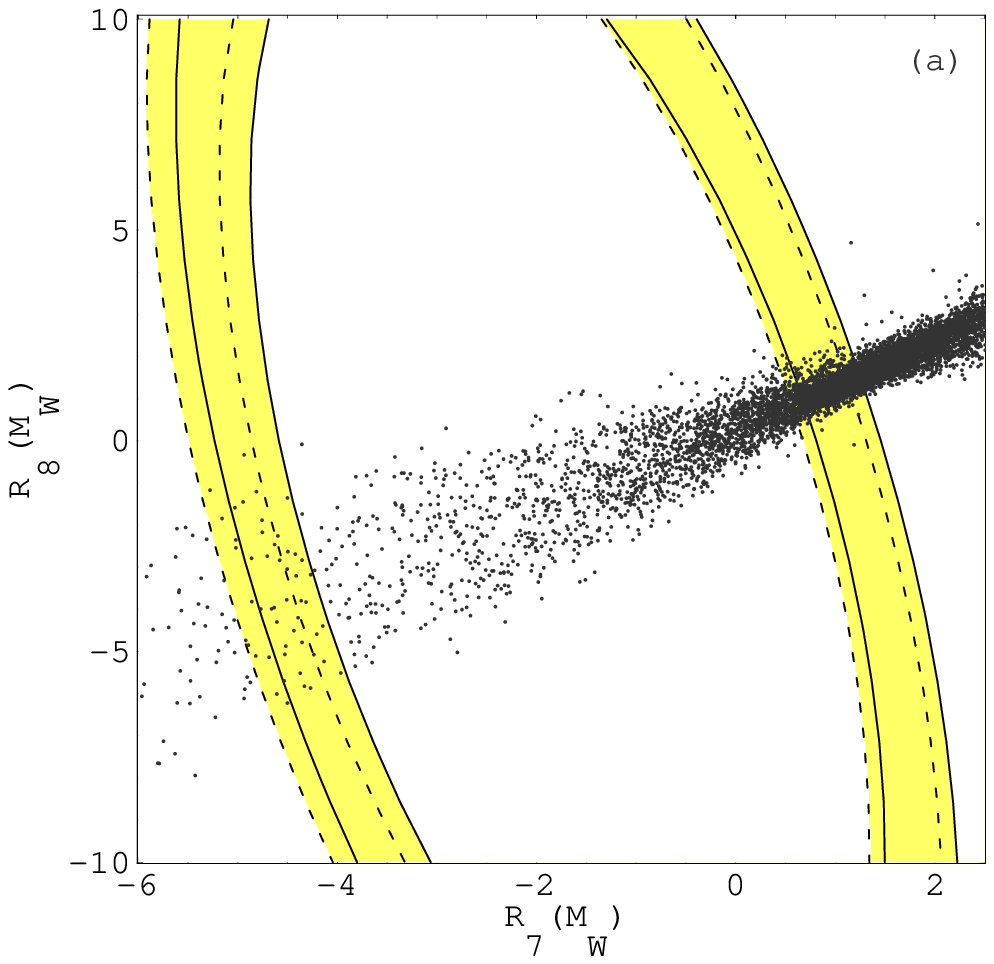,width=0.37\linewidth}
\hspace*{1cm}
\epsfig{file=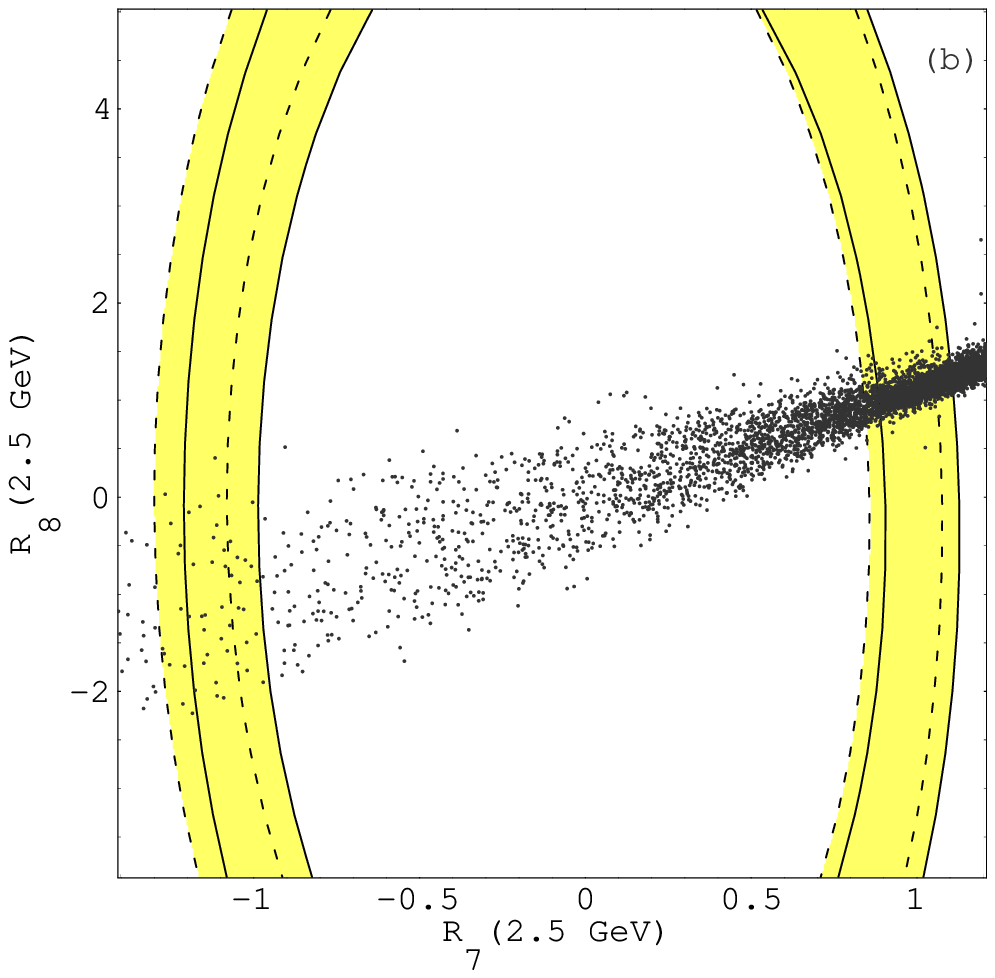,width=0.37\linewidth}
\caption{\it $\cl{90}$ bounds in the $[R_7 (\m), R_8(\m)]$ plane
following from the world average $B\to X_s \g$ branching ratio for
$\mu=m_W$ (left-hand plot) and $\mu=2.5$ GeV (right-hand plot).
Theoretical uncertainties are taken into account. The solid and dashed
lines correspond to the $m_c = m_{c,pole}$ and $m_c =
m_{c}^{\overline{MS}} (\m_b)$ cases respectively. The scatter points
correspond to the expectation in MFV models (the ranges of the SUSY
parameters are specified in the text).}
\label{fig:bsg}
\end{center}
\end{figure}
In our analysis, we make the assumption that the dominant new physics
effects can be implemented by using the SM operator basis for the
effective hamiltonian. The Wilson coefficients that are constrained by
the set of measurements given in Eqs.~(\ref{bsgexp})--(\ref{bseeexp})
are $C_{7,8,9,10}$. Note that $b\to s \ell^+ \ell^-$ transitions
depend on $C_7$ and $C_8$ only through the effective coefficient $A_7
(C_7,C_8)$~\cite{aghl}. Our first step consists in writing $\branch
(B\to X_s \g)$ as a function of $R_{7,8}(\mu)\equiv A_{7,8}^{\rm tot}
(\mu) / A_{7,8}^{\rm SM} (\mu)$ (following \rf{Kagan:1999ym}) and
plotting the regions in the $[R_7,R_8]$ plane that are allowed by
Eq.~(\ref{bsgexp}). Requiring $|R_8(\mu_W)| \leq 10$ in order to
satisfy the constraints from the decays $B\to X_s g$ and $B\to X_{c
\!\! /}$ (where $X_{c\!\! /}$ denotes any hadronic charmless final
state) we are, then, able to extract the bounds on $A_7 (\m_b)$. It
was recently pointed out in \rf{Gambino:2001ew} that the charm mass
dependence of the $B\to X_s \g$ branching ratio was underestimate in
all the previous analyses. Indeed, the replacement of the pole mass
($m_{c,pole}/m_{b,pole} = 0.29 \pm 0.02$) with the $\overline{MS}$
running mass ($m_{c}^{\overline{MS}} (\m_b)/m_{b,pole} = 0.22 \pm
0.04$) increases the branching ratio of about $11 \%$. In order to
take into account this additional source of uncertainty, we work out
the constraints on the Wilson coefficients for both choices of the
charm mass; we will then use the loosest bounds in the $b\to s \ell^+
\ell^-$ analysis. We present the resulting allowed regions in
Figs.~\ref{fig:bsg}a and \ref{fig:bsg}b (note that according to the
previous discussion we are interested in the bounds at $\mu_b = 2.5 \;
\gev$).  The regions in \fig{fig:bsg}b translate in the following
allowed constraints:
\bea
\cases{
m_c/m_b = 0.29:  \;\;\; A_7^{\rm tot} (2.5 \; \gev)\in [-0.37,-0.18] \; \& \; [0.21,0.40] \; , & \cr 
m_c/m_b = 0.22:  \;\;\; A_7^{\rm tot} (2.5 \; \gev)\in [-0.35,-0.17] \; \& \; [0.25,0.43] \; . & \cr}
\eea
In the subsequent numerical analysis we impose the union of the above
allowed ranges
\bea
\label{a7lim}
 -0.37 \leq A_7^{\rm tot,<0} (2.5 \; \gev) \leq -0.17 
  & \& &
  0.21 \leq A_7^{\rm tot,>0} (2.5 \; \gev) \leq 0.43 
\eea
calling them $A_7^{\rm tot}$--positive and $A_7^{\rm tot}$--negative
solutions.

\begin{figure}
\begin{center}
\epsfig{file=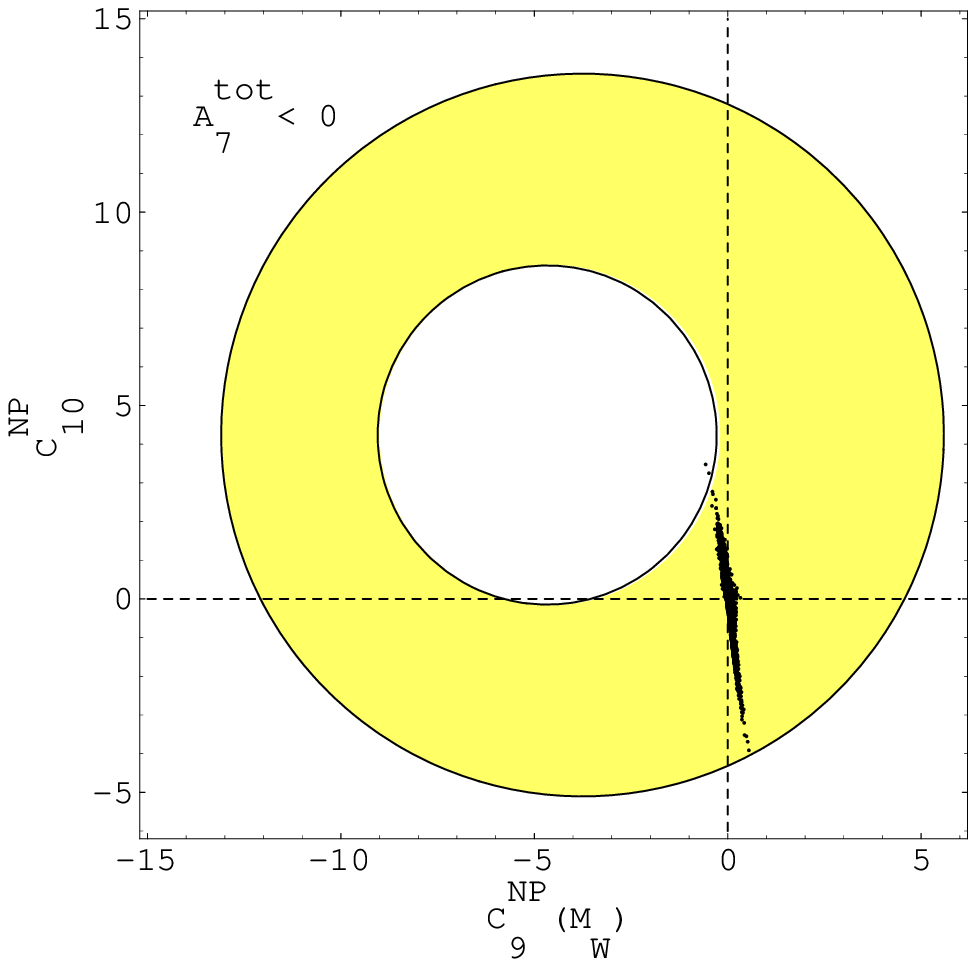,width=0.37\linewidth}
\hspace*{1cm} 
\epsfig{file=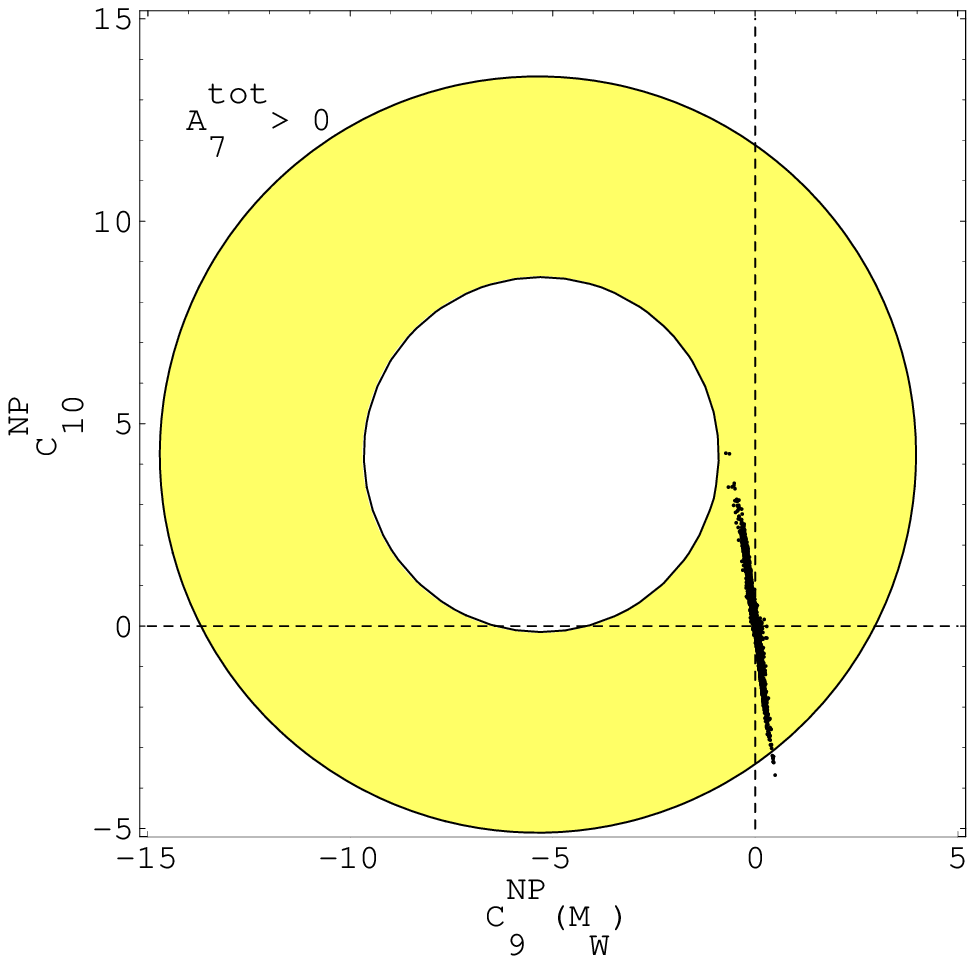,width=0.37\linewidth}
\caption{\it {\bf NNLO Case.} Superposition of all the
constraints. The plots correspond to the $A_7^{\rm tot}(2.5 \;
\gev)<0$ and $A_7^{\rm tot}(2.5 \; \gev) >0$ case, respectively. The
points are obtained by means of a scanning over the EMFV parameter
space and requiring the experimental bound from $B\to X_s \g$ to be
satisfied. \hspace*{5.5cm}}
\label{fig:total}
\end{center}
\end{figure}

The bounds in the $[C_9^{NP} (\mu_W),C_{10}^{NP}]$ plane, implied by
the experimental results (\ref{bkmmexp})--(\ref{bseeexp}) are
summarized in Fig.~\ref{fig:total}, where the two plots correspond
respectively to the $A_7^{\rm tot}$--positive and $A_7^{\rm
tot}$--negative solutions just discussed. Note that the overall
allowed region is driven by the constraints emanating from the decays
$B\to X_s e^+ e^-$ (outer contours) and $B\to K \mu^+ \mu^-$ (inner
contours).
\section{Expectation in supersymmetry}
In this section we analyze the impact of the $b\to s\g$ and $b\to s
\ell^+ \ell^-$ experimental constraints on several supersymmetric
models. We will first discuss the more restricted framework of the
minimal flavour violating MSSM, and then extend the analysis to more
general models in which new SUSY flavour changing couplings are
allowed.
\subsection{Minimal flavour violation}
As already known from the existing literature (see for instance
\rf{LMSS}), minimal flavour violating (MFV) contributions are
generally too small to produce sizable effects on the Wilson
coefficients $C_9$ and $C_{10}$. In the MFV scheme all the genuine new
sources of flavour changing transitions other than the CKM matrix are
switched off, and the low energy theory depends only on the following
parameters: $\m$, $M_2$, $\tan \b$, $M_{H^\pm}$, $M_{\tilde t_2}$ and
$\theta_{\tilde t}$ (see Ref.~\cite{aghl} for a precise definition of
the various quantities). Scanning over this parameter space and taking
into account the lower bounds on the sparticle masses as well as the
$b\to s \g$ constraint given in \eq{bsgexp}, we derive the ranges for
the new physics contributions to $C_9$ and $C_{10}$. In order to
produce bounds that can be compared with the model independent allowed
regions plotted in \fig{fig:total}, we divided the surviving SUSY
points in two sets, according to the sign of $A_7^{\rm tot}$:
\bea
A_7^{\rm tot}<0 \Rightarrow \cases{C_9^{MFV}(\m_W) \in [-0.2, 0.4]\, , \cr  
                                   C_{10}^{MFV}    \in [-1.0, 0.7]}\, . \\
A_7^{\rm tot}>0 \Rightarrow \cases{C_9^{MFV}(\m_W) \in [-0.2, 0.3]\, , \cr  
                                   C_{10}^{MFV}    \in [-0.8, 0.5] \; .}
\eea
We stress that the above discussion applies to any supersymmetric model
with flavour universal soft-breaking terms, such as
minimal supergravity MSSM and gauge-mediated supersymmetry
breaking models. Beyond-the-SM flavour violations in such models are
induced only via renormalization group running, and are tiny.
Hence, they can be described by MFV models discussed in this paper.

Before finishing this subsection and starting our discussion on models
with new flavour changing interactions, let us show in more detail the
impact of $b\to s \gamma $ on MFV models. The scatter plot presented
in Fig.~\ref{fig:bsg} is obtained varying the various MFV SUSY
parameters and shows the strong correlation between the values of the
Wilson coefficients $C_7$ and $C_8$. In fact, the SUSY contributions
to the magnetic and chromo--magnetic coefficients differ only because
of colour factors and loop-functions.

\subsection{Chargino contributions: Extended--MFV models}
A basically different scenario arises if chargino--mediated penguin
and box diagrams are considered in connection with non--vanishing mass
insertions (See \rf{mia} for a definition of the so--called mass
insertion approximation (MIA)). As can be inferred by Table~4 in
\rf{LMSS}, the presence of a light $\tilde t_2$ generally gives rise
to large contributions to $C_9$ and especially to $C_{10}$. In the
following, we will concentrate on the so--called Extended MFV (EMFV)
models introduced in \rf{AL}. In these models we can fully exploit the
impact of chargino penguins with a light $\tilde t$ still working with
a limited number of free parameters.  EMFV models are based on the
heavy squarks and gluino assumption.  In this framework, the charged
Higgs and the lightest chargino and stop masses are allowed to be
light while the rest of the SUSY spectrum is assumed to be degenerate
and heavier than $1 \; \tev$. The assumption of a heavy gluino
suppresses any possible gluino--mediated SUSY contribution to low
energy observables. Note that even in the presence of a light gluino
these penguin diagrams remain suppressed due to the heavy down squarks
present in the loop. In the MIA approach, a diagram can contribute
sizably only if the inserted mass insertions involve the light
stop. This leaves us with only two unsuppressed flavour changing
sources other than the CKM matrix, namely the mixings $\tilde u_L -
\tilde t_2$ (denoted by $\d_{\tilde u_L \tilde t_2}$) and $\tilde c_L
- \tilde t_2$ (denoted by $\d_{\tilde c_L \tilde t_2}$). The
phenomenological impact of $\delta_{\tilde t_2 \tilde u_L}$ has been
studied in \rf{AL} and its impact on the $b\to s \g$ and $b\to s
\ell^+ \ell^-$ transitions is indeed negligible. Therefore, we are
left with the MIA parameter $\delta_{\tilde t_2 \tilde c_L}$ only.
Scanning over the SUSY parameter space ($\mu$, $M_2$, $\tan \beta$,
$M_{\tilde t_2}$, $\sin \theta_{\tilde t}$, $M_{H^\pm}$, $M_{\tilde
\nu}$ and $\delta_{\tilde t_2 \tilde c_L}$) and imposing the
constraints from the sparticle masses lower bounds and $B\to X_s \g$,
we obtain the points plotted in \fig{fig:total}. Note that these SUSY
models can account only for a small part of the region allowed by the
model independent analysis of current data.
\section{Conclusions}
We have presented theoretical branching ratios, model independent
analyses and SUSY predictions for the rare $B$ decays $B \to X_s
\ell^+ \ell^-$ and $B \to (K,K^*) \ell^+ \ell^-$, incorporating the
NNLO improvements. The dilepton invariant mass spectrum is calculated
in the NNLO precision in the low dilepton invariant mass region,
$\hat{s} < 0.25$ and an extrapolation is used for $\hat{s} >
0.25$. Current B factory experiments will soon probe these decays at
the level of the SM sensitivity and we stress the need to measure the
inclusive decays $B \to X_s \ell^+ \ell^-$ in the low dilepton mass
range for a proper comparison with the SM expectations.
\section*{Acknowledgments}
I am much indebted to A. Ali, C. Greub and G. Hiller for a most
enjoyable collaboration. It is a pleasure to thank the organizers of
the XXXVIIth Rencontres de Moriond for the very sparkling and snowy
atmosphere. I acknowledge financial support from the Alexander von
Humboldt Foundation.
\section*{References}

\end{document}